\documentclass[12pt]{iopart}
\eqnobysec
\def\eqn{\begin{equation}}
\def\enq{\end{equation}}
\def\ena{\begin{array}}
\def\eqa{\end{array}}
\def\cn{{\cal N}}
\newfont{\Bbb}{msbm10 scaled 1200}     
\newcommand{\mathbb}[1]{\mbox{\Bbb #1}}
\def\IZ{{\mathbb Z}}
\def\IR{{\mathbb R}}
\begin{document}
{\vbox{
\rightline{RUNHETC-99-42}
\rightline{hep-th/9911147}}}
\title{A brief review of ``little string theories''}
\author{Ofer Aharony}
\address{
Department of Physics and Astronomy\\
Rutgers University\\
Piscataway, NJ 08855--0849, USA}
\begin{abstract}

This is a brief review of the current state of knowledge on ``little
string theories'', which are non-gravitational theories having several
string-like properties. We focus on the six dimensional maximally
supersymmetric ``little string theories'' and describe their
definition, some of their simple properties, the motivations for
studying them, the DLCQ and holographic constructions of these
theories and their behaviour at finite energy density.
(Contribution to the proceedings of Strings '99 in Potsdam, Germany.)
\end{abstract}
\section{Introduction}

One of the most surprising results which came out of the developments
in string theory in the last few years is the existence of consistent
non-gravitational theories in five and six space-time dimensions, even
though no consistent Lagrangians are known for interacting theories in
these dimensions. Generally these theories have been discovered by
considering some limit of string theory configurations involving
5-branes and/or singularities. The higher dimensional
non-gravitational theories may be divided into two classes. One class,
which generally arises from a low-energy limit of string theory (or M
theory), includes superconformal field theories in five and six
dimensions\footnote{Supersymmetry plays an essential role in proving
the existence of these theories, it is not clear if non-supersymmetric
theories also exist above four dimensions or not.}. These theories
seem to be standard local field theories, even though they have no
good Lagrangian description (they are sometimes called ``tensionless
string theories'' since many of them have BPS-saturated strings on
their moduli space whose tensions go to zero at the conformal
point). We will not discuss these theories here. The other class of
theories, which was given the name ``little string theories'' (LSTs)
by \cite{lms}, is generally obtained by taking the string coupling to
zero in some configuration of NS 5-branes and/or singularities which
is not well-described by perturbation theory, and which in fact
remains non-trivial (but decoupled from gravity) even after taking the
string coupling to zero. The string scale $M_s$ remains constant in
this limit and plays an important role in the dynamics of these
theories; they appear to be non-local theories with some string
theory-like properties. In this contribution I will try to summarize
all that is currently known about the simplest theories of this type,
which are six dimensional theories with 16 supercharges arising as the
$g_s \to 0$ limit of NS 5-branes in type IIA or type IIB string
theory. There are many interesting results on lower dimensional
LSTs, LSTs with less supersymmetry and compactifications of LSTs, that
I will not have room to review here. The transparencies and audio for
this talk are available at \cite{trans}.

\subsection{Definition and simple properties of ``little string theories''}

The simplest definition of six dimensional LSTs with 16 supercharges
comes from looking at $k$ parallel and overlapping NS 5-branes in type
IIA or type IIB string theory, with $k > 1$, and taking the string
coupling $g_s \to 0$ \cite{natistr} (see also \cite{brs,dvv}). The
gravitational interactions go to zero in this limit, but the couplings
of the fields associated with the NS 5-branes remain non-trivial even
after taking $g_s$ to zero; this can be seen, for instance, by
analyzing the low-energy theory (as described below). The string scale
$M_s$ is kept finite in the limit, and it is the only parameter of the
LSTs (except for the discrete parameter $k$); in particular, there is
no continuous dimensionless coupling parameter in these theories, so
unlike conventional string theories they have no obvious weakly
coupled limits which can serve as the starting point for a
perturbative expansion. The NS 5-branes break half of the
supersymmetry of type II string theories, and thus there are no forces
between them and we can indeed put $k$ of them on top of each
other. In six dimensions (like in ten or two dimensions) supersymmetry
is generally chiral, and the dimensional reduction of type II string
theory to six dimensions has $\cn=(2,2)$ supersymmetry. In the type
IIA case the NS 5-branes preserve a chiral half of the supersymmetry
so the resulting LST has $\cn=(2,0)$ supersymmetry, while in the type
IIB case it has $\cn=(1,1)$ supersymmetry.

We can obtain equivalent definitions of the LSTs by using various
duality symmetries. Using the duality between type IIA string theory
and M theory we see that the $(2,0)$ LST can also be derived from $k$
5-branes in M theory, with a transverse circle of radius $R$, in the
limit $R \to 0$, $M_p \to \infty$ with $RM_p^3 = M_s^2$ kept
constant. Using S-duality in type IIB we see that the $(1,1)$ LST can
be derived from the $g_s \to \infty$ limit of $k$ D5-branes in type
IIB string theory. 

Additional definitions arise from recalling that $k$ NS 5-branes with
a transverse circle are T-dual to an $A_{k-1}$ singularity with a
compact circle \cite{ov}, and noting that the existence of such a
circle does not affect the decoupled theory on the NS 5-branes (it
affects only the bulk modes which decouple when $g_s \to 0$). Thus,
by using this T-duality, we conclude that the $(2,0)$ LSTs arise also
as the $g_s \to 0$ limit of type IIB string theory on an $A_{k-1}$
singularity, and the $(1,1)$ LSTs arise also as the $g_s \to 0$ limit
of type IIA string theory on an $A_{k-1}$ singularity\footnote{There
is a subtlety here involving the degree of freedom corresponding to
the center of mass position of the NS 5-branes which is not evident in
the T-dual picture, generally we can ignore this degree of freedom
since it is free and decoupled.}. The last definition suggests a
generalization to the $g_s \to 0$ limit of $D_n$ and $E_n$ type
singularities, which also correspond to LSTs with 16 supercharges; we
will not discuss these theories in detail here, but let us note that
the full classification of LSTs in six dimensions with 16 supercharges
includes theories with $(2,0)$ and $(1,1)$ supersymmetry of type $G$
(where $G=A_k,D_k$ or $E_k$) for any simply-laced Lie algebra $G$.

Some simple properties of the LSTs with 16 supercharges follow from
the above definitions :
\begin{enumerate}
\item
Low-energy behaviour : For the $(2,0)$ LSTs it follows from the M
theory definition that the low-energy behaviour (well below the
characteristic scale $M_s$) is given by the low-energy theory on $k$
M5-branes, which is an $\cn=(2,0)$ superconformal theory (see,
e.g., \cite{natisixteen}). For the $(1,1)$ LSTs it follows from the
definition using D5-branes in type IIB string theory that the
low-energy behaviour is given by an $\cn=(1,1)$ $U(k)$ gauge theory
whose gauge coupling is $g_{YM}^2 = 1 / M_s^2$. For large $k$, 't Hooft
scaling suggests that the perturbative gauge theory (which is free at
low energies) breaks down at an
energy squared scale of order $1 / g_{YM}^2 k = M_s^2 / k$.
\item
The moduli space metric of theories with 16 supercharges cannot
receive any quantum corrections. For the $(1,1)$ LSTs the type IIB
constructions show that it is $\IR^{4k}/S_k$, corresponding to the
transverse positions of the $k$ identical 5-branes. For the $(2,0)$
LSTs the M theory construction shows that it is $(\IR^4 \times
S^1)^k/S_k$, where the radius of the $S^1$ is $M_s^2$ (recall that the
canonical dimension of a scalar field in six dimensions is two). The
low-energy theory at generic points in the moduli space involves $k$
tensor multiplets in the $(2,0)$ case and $k$ vector multiplets in the
$(1,1)$ case.
\item
Type IIA string theory on a circle of radius $R$ is T-dual to type IIB
string theory on a circle of radius $1/M_s^2 R$, and an NS 5-brane
wrapped on the circle is transformed under this duality to an NS
5-brane wrapped on the dual circle. Thus, T-duality commutes with the
limit defining the LSTs, and the $(2,0)$ $A_{k-1}$ LST compactified on
a circle of radius $R$ is dual to the type $(1,1)$ $A_{k-1}$ LST
compactified on a circle of radius $1/M_s^2 R$. Similarly, the LSTs
compactified on $T^d$ have an $O(d,d,\IZ)$ T-duality symmetry. Note
that this shows that T-duality can exist even in non-gravitational
theories. The existence of such a T-duality symmetry is the first
indication we see that the LSTs are non-local; in particular, after
toroidal compactification, they do not have a unique energy-momentum
tensor (the same theory can be coupled to different gravitational
backgrounds).
\item
All LSTs have BPS-saturated strings of tension $T = M_s^2$ (at the
origin of moduli space; in some cases there are more BPS-saturated
strings away from the origin). These may be viewed as marginally bound
states of fundamental strings with the NS 5-branes. At the origin of
moduli space, the $(2,0)$ LST has no other BPS states, while in the
$(1,1)$ LST the low-energy gluons (and their superpartners) are
massless BPS particles. After compactification there are many
additional BPS states which we will not discuss here (see, e.g.,
\cite{natistr}).
\end{enumerate}

\subsection{Motivations}

At first sight there is no reason to be interested in ``little string
theories'', since they are neither directly relevant for studying
quantum field theories (being non-local theories) nor for studying
quantum gravity (being non-gravitational theories). However, the very
fact that these theories are, in some sense, intermediate between
local field theories and ``standard'' string theories means that they
may be able to teach us about both. For example, the fact that LSTs
have some string theory-like properties, like T-duality and a Hagedorn
spectrum (to be discussed below), means that such properties can
appear also in non-gravitational theories, and it may be easier to
understand them in that context. The existence of non-gravitational
Lorentz-invariant theories which are intrinsically non-local is very
interesting in itself, and it would be nice to know how to define
these theories and what is the proper way to think about them.

Some more concrete motivations for studying these theories are :
\begin{enumerate}
\item
The original construction of LSTs was motivated by the fact that they
(or rather their compactification on $T^5$) arise as the discrete
light-cone quantization (DLCQ) of M theory on $T^5$ \cite{brs} with
$k$ units of longitudinal momentum.
\item
Compactifications of the LSTs lead to many interesting local and
non-local field theories. For example, the low-energy limit of the
LSTs on $T^2$ gives four dimensional $\cn=4$ SYM theories, and the
compactification of LSTs on other manifolds gives theories related to
QCD \cite{wittenqcd}.
\item
As we will see in the next section, the LSTs are a rare example of a
theory whose discrete light-cone quantization (DLCQ) is simple, and we
can use it as a toy model to learn about DLCQ.
\item
As we will see in section 3, linear dilaton backgrounds of string
theory seem to be holographically dual to LSTs, and thus studying LSTs
can teach us about holography in linear dilaton backgrounds (and
perhaps give us clues about how to define holography in general
backgrounds).
\end{enumerate}

\section{DLCQ constructions}

The definitions of LSTs given above have not been useful so far for
making any explicit computations in these theories, so one would like
to have more direct definitions of the LSTs. The first such definition
was found using discrete light-cone quantization (DLCQ), which is a
quantization of the theory compactified on a light-like circle of
radius $R$ with $N$ units of momentum around the compact circle. In
the large $N$ limit the momentum around the circle becomes effectively
continuous, and it is believed that all six dimensional results may be
reproduced. The advantage of DLCQ comes from the fact that
negative-momentum modes decouple in the light-cone frame, so the
dynamics involves just the positive-momentum modes (for example it can
only involve a finite number of particles). To derive a DLCQ theory we
need to exactly integrate out the zero modes of fields (carrying no
momentum in the compact direction), which is usually
complicated. However, there is a small class of theories where one can
derive the exact DLCQ theory, since they have enough supersymmetry to
determine the Lagrangian after integrating out the zero modes, and the
six dimensional LSTs fall into this class.

The DLCQ of LSTs may be derived either by following Seiberg's
prescription of regarding the DLCQ as the limit of a compactification
on a small space-like circle \cite{natimat}, or by starting from
M(atrix) theory, which is the DLCQ of string theory or M theory, in
configurations with 5-branes or singularities, and taking the limit
defining the LSTs. All the derivations lead to the same result and
they are detailed in the literature (and reviewed in \cite{ab}), we
will present here only the result.

The DLCQ of the $A_{k-1}$ $(2,0)$ LST with $N$ units of momentum is
\cite{abkss,edhiggs} the $1+1$ dimensional $\cn=(4,4)$ supersymmetric
sigma model on the $4Nk$-dimensional moduli space of $N$ instantons in
$SU(k)$ (on $\IR^4$), compactified on a circle of radius $\Sigma = 1 /
R M_s^2$. Equivalently (by the ADHM construction), it is the conformal
theory describing the low-energy limit of the Higgs branch of the
$\cn=(4,4)$ $U(N)$ SQCD theory with an adjoint hypermultiplet and $k$
hypermultiplets in the fundamental representation. The moduli space of
instantons is singular, but the sigma model on it seems to make
sense. The spectrum of chiral operators of the $(2,0)$ LST may be
computed in this description (a similar computation in the $0+1$
dimensional sigma model on the moduli space of instantons was
performed in \cite{abs}).

For the $A_{k-1}$ $(1,1)$ LSTs it turns out to be complicated to write
down the usual DLCQ, but simple to write down a DLCQ with a Wilson
line of the low-energy $U(k)$ gauge group around the light-like circle
\cite{sethi,gs,ab}. In the large $N$ limit the effect of the Wilson
line is expected to disappear. The DLCQ with the Wilson line is the
conformal theory which arises as the low-energy limit of the Coulomb
branch of the $1+1$ dimensional $\cn=(4,4)$ $U(N)^k$ gauge theory with
bifundamental hypermultiplets for adjacent groups (when we arrange the
$k$ gauge groups in a circle, as in a quiver diagram). Again, the
conformal theory is compactified on a circle of radius $\Sigma = 1 /
M_s^2 R$. Removing the Wilson line corresponds to taking some of the
gauge couplings to infinity before taking the low-energy limit.

The DLCQ gives us an explicit non-perturbative definition of the
``little string theories'', enabling us in principle to compute all
their states and correlation functions. Unfortunately, in practice
such computations are extremely difficult; the conformal theories
involved are very complicated even before we take the large $N$ limit,
and we have to take this limit to obtain results about the LSTs in six
dimensions. Thus, no dynamical computations have been made so far
using the DLCQ, but only identifications of some operators and states
(see, e.g., \cite{abs,ab}). As usual in DLCQ, it is quite complicated
to study other points on the moduli space of the LSTs or
compactifications, both of which lead to quite different DLCQ
theories. For example, the DLCQ of the $(2,0)$ $A_{k-1}$ theory at a
generic point on its moduli space is given by the deformation of the
Higgs branch SCFT described above by generic masses for the $k$
hypermultiplets in the fundamental representation (which is a relevant
deformation of the Higgs branch SCFT).

\section{Holographic constructions}

The second useful construction of LSTs is a holographic dual along the
lines of the AdS/CFT correspondence; unlike the previous construction
this one cannot serve as a definition of the LSTs since we have no
non-perturbative definition of the string/M theory backgrounds that
are involved, but it seems to be more useful in computing correlation
functions in the LSTs (at least at low energies). The AdS/CFT
correspondence \cite{juan,gkp,wittenads,magoo} states that local
conformal field theories are often dual to string/M theory
compactifications including AdS spaces, and it was generalized to some
other classes of local field theories as well (though these are often
dual to spaces with regions of high curvature). However, there is no
reason for general backgrounds of string/M theory (or any other theory
of quantum gravity) to be holographically dual to a local field
theory, and it seems likely that the theory which is holographically
dual to quantum gravity in Minkowski space is highly non-local. LSTs
provide the first example of holography for a non-local
non-gravitational theory, and it turns out that generally LST-like
theories are holographically dual to backgrounds which asymptote to
string theory in a linear dilaton background \cite{abks}.

As in the AdS/CFT correspondence, the space which is holographically
dual to LSTs may be derived by starting from the string theory
background corresponding to NS 5-branes and taking the limit of $g_s
\to 0$ (where $g_s$ is the asymptotic string coupling) which defines
the LSTs \cite{abks,imsy,bst}. For $(2,0)$ LSTs the correct procedure
is actually to start from the background corresponding to M5-branes
with a transverse circle, as described above, since a background of NS
5-branes in string theory does not correspond to a configuration which
is localized in the $S^1$ coordinates of the moduli space. Starting
with such a background and taking the appropriate limit leads to the
space 
\eqn
\label{holo}
l_p^{-2} ds^2 = H^{-1/3} dx_6^2 + H^{2/3} (dx_{11}^2 + dU^2 + U^2
d\Omega_3^2),
\enq
where $dx_6^2$ is the metric on $\IR^6$, $d\Omega_3^2$ is the metric
on $S^3$, $x_{11}$ is compactified on a circle with radius $M_s^2$,
\eqn
H = \sum_{j=-\infty}^{\infty} {{\pi k} \over {(U^2 + (x_{11} - 2\pi j
M_s^2)^2)^{3/2}}},
\enq
and there are also $k$ units of 4-form flux on $S^3 \times S^1$. M
theory compactified on this space is holographically dual to the
$A_{k-1}$ $(2,0)$ LST.

The space (\ref{holo}) is quite complicated, but it simplifies in the
asymptotic regions of space. For small $U$ and $x_{11}$ the space
(\ref{holo}) becomes just $AdS_7\times S^4$, which is holographically
dual to the $(2,0)$ SCFTs, as expected since these SCFTs are the
low-energy limit of the $(2,0)$ LSTs. For large $U$ the physical
radius of the $x_{11}$ circle becomes very small and it is more
appropriate to view the background as a type IIA string theory
compactification. Defining a new variable $\phi$ by $l_s^2 U =
\sqrt{k} e^{\phi/\sqrt{k} l_s}$, the string metric is simply
\eqn
\label{holostring}
ds_{string}^2 = dx_6^2 + d\phi^2 + kl_s^2 d\Omega_3^2,
\enq
with a linear dilaton in the $\phi$ direction,
\eqn
g_s(\phi) = e^{-\phi/\sqrt{k} l_s},
\enq
and $k$ units of 3-form flux on the $S^3$.
Note that for large $k$ the curvatures are small everywhere, so
supergravity is a good approximation at low energies.
Even though the behaviour of this space near the boundary is very
different from that of AdS space, it seems that the same principles
apply also in this case; for example, as in AdS/CFT, correlation
functions in the LSTs are identified with the response of string/M
theory on the background (\ref{holo}) to turning on boundary values for
non-normalizable modes of the fields. In the space (\ref{holo}) some of
these modes are just the incoming and outgoing waves in the $\phi$
direction, so that some correlation functions of the LST correspond to
the S-matrix for scattering in the $\phi$ direction.

Similarly, the type $(1,1)$ LSTs are holographically dual to the
near-horizon limit of type IIB NS 5-branes; unfortunately this
background becomes singular for small $U$ (there is no analog of the M
theory region in (\ref{holo})) so this description is less useful for
computations of correlation functions, but it can still be used for
identifying the operators and some states of these LSTs as described
below.

The holographic description of the LSTs is useful for :
\begin{enumerate}
\item
As in the AdS/CFT correspondence, some correlation functions may be
computed using supergravity. In this case the supergravity
approximation is valid for the $(2,0)$ LSTs at large $k$ and at
energies well below the string scale $M_s$, where stringy corrections
in the background (\ref{holostring}) become important. The computation
of the 2-point function of the energy-momentum tensor in the LSTs was
described in \cite{shiraz}, and it is possible to compute also other
correlation functions (though the actual computations are quite
complicated). In particular, supergravity gives an exact description
of the analog of the 't Hooft limit for the $(2,0)$ LSTs, in which
$M_s$ and $k$ are taken to infinity with the scale $M_s^2 / k$ (which
in the $(1,1)$ case is the inverse 't Hooft gauge coupling) kept
constant.
\item
A big difference between the background (\ref{holo}) and backgrounds
which are dual to other field theories is that near the boundary of
(\ref{holo}) the string coupling goes to zero and the curvatures are
small, so one can compute the spectrum of fields exactly (and not just
for large $k$ as in other cases). The full spectrum of chiral fields
in the LSTs was computed in this way in \cite{abks}, and turned out to
be exactly the same for all values of $k$ (in the $A_{k-1}$ case) as
the spectrum of chiral fields in the field theories which arise as the
low-energy limit of the LSTs.
\item
The holographic description can be used to reliably compute some of
the states in the LSTs, which are states propagating in the weakly
coupled region of (\ref{holo}). For example, one has the states in the
supergraviton multiplet propagating with some momentum in the $\phi$
direction, and these look like a continuum of states from the six
dimensional point of view (the mass shell condition relates the
6-dimensional momentum to the $\phi$-momentum but it does not
determine its magnitude). It turns out that for the states in the
supergravity multiplet this continuum of states starts at a scale of
order $M^2 \simeq M_s^2/k$, while other string states in the weak
coupling region also give rise to continuous spectra in the LST,
starting at higher values of $M^2$. The implications of the existence
of this continuum are not clear. The same continuum of states can also
be seen in the DLCQ formalism \cite{ab}, where it is related to the
continuum of states found in sigma models on orbifold singularities
with zero theta angle.
\item
The holographic description can be used to analyze the behaviour at
finite temperature and energy density, as discussed in the next section.
\end{enumerate}

\section{Behaviour at finite energy density}

We have already seen some indications that LSTs are not local field
theories, but the strongest indication for this seems to come from
analyzing the equation of state of the LSTs at finite temperature or
energy density. For local field theories the high-energy behaviour of
the density of states is always an exponential of a power of the
energy density which is less than one, while for the LSTs we will see
that it is exponential in the energy density.

The most reliable way to compute the equation of state of the LSTs is
to use their holographic description. Holographic dualities relate
finite temperature states of non-gravitational theories to black hole
configurations, with the Hawking temperature of the black hole equated
with the field theory temperature, and the field theory energy equated
with the mass of the black hole. The entropy of such black holes can
be computed by the usual Bekenstein-Hawking formula which relates it
to their area. In the case of LSTs the appropriate black holes (at
large enough energy densities) are the near-horizon limits of
near-extremal NS5-branes and this computation was first done in
this context in \cite{malstro}. The result is
\eqn
\label{eos}
E = T_H S; \quad T_H \simeq M_s / \sqrt{6k},
\enq
and it is reliable when the curvatures in the black hole background
are small (requiring $k \gg 1$) and when the string coupling at the
black hole horizon is small (recall that the string coupling becomes
weaker and weaker as one goes towards the boundary), requiring that
the energy density $\mu \equiv E / V$ satisfies $\mu \gg k M_s^6$. Thus, we
find that for large $k$ and large energy densities the equation of
state is Hagedorn-like, corresponding to a limiting temperature in
the theory of $T_H \simeq M_s / \sqrt{6k}$; above some energy density
which is smaller than $k M_s^6$ the specific heat diverges and 
increasing the energy density no
longer increases the temperature, which remains $T = T_H$. 

This behaviour is similar to the behaviour of single-string states in
free string theory, where $T_H \simeq M_s$; however, in string theory
it is believed (at least in some cases) that at the temperature $T_H$
there is a phase transition to a different phase where a state
becoming massless at $T=T_H$ condenses \cite{atick}, while in the LSTs
there is no evidence for any states becoming massless at $T=T_H$, so
it seems that $T_H$ is really a limiting temperature (and the
canonical ensemble cannot be defined beyond this temperature).

It would be nice to reproduce the behaviour (\ref{eos}) also in the DLCQ
description of the LSTs, but this seems rather difficult. The behaviour
(\ref{eos}) is, in fact, reproduced by a naive counting of the density of
states in the DLCQ description \cite{ahaban}, but the relevant states
are really only those whose energy scales as $1/N$ in the large $N$
limit, and identifying all these states is much more difficult. Some
states whose energy scales as $1/N$ were identified in \cite{ab}, and
they correspond to a density of states in space-time which is
exponential but with a smaller exponent corresponding to $\tilde{T}_H
\simeq M_s/\sqrt{12}$; it would be interesting to identify in the DLCQ
the other states which give rise to (\ref{eos}).

It is not clear how to interpret the exponential behaviour of the
equation of state in the LSTs. The fact that the density of states
depends only on the energy and not on the volume suggests that
generic high-energy states are single-object configurations, perhaps
similar to long, space-filling strings. It is not yet known at which
energy density the behaviour (\ref{eos}) begins, and what is the equation
of state in the regime $\mu < M_s^6$ (which is relevant, in
particular, for the 't Hooft limit of the LSTs described above).

General arguments (with some assumptions about the behaviour of
generic operators in the LSTs) suggest that the behaviour (\ref{eos})
implies the non-existence of correlation functions of operators at
separations smaller than $T_H^{-1}$. Presumably, there are no local
operators in these theories, but only operators ``smeared'' on
distance scales of at least $T_H^{-1}$. Computations in the
holographic formulation give correlation functions of operators (which
are naively local) in momentum space, but it seems that the Fourier
transform of the results to position space does not exist for small
separations \cite{peet}, consistent with this non-locality. Note that
the scale of non-locality suggested by these arguments is different
(by a factor of order $\sqrt{k}$) from the scale of non-locality
suggested by the T-duality symmetry.

\section{Future directions}

Much progress has been made in the last few years in understanding
``little string theories'', but many open questions remain. Some
interesting open questions are :
\begin{enumerate}
\item
The behaviour of LSTs at high energies is clearly not governed by a
field theoretical fixed point, and it is interesting to understand
what the high-energy behaviour is. There are some indications that the
theory becomes weakly coupled at high energies, such as the fact that
in the holographic description the string coupling vanishes near the
boundary, and that the string coupling is small everywhere in the
holographic dual of the theory at high energy densities. It would be
interesting to understand if there is some sense in which the theory
becomes weakly coupled at high energies (perhaps as in asymptotically
free field theories) and if there is any simple description of the
high-energy limit.
\item
The behaviour at intermediate energy scales, such as those governed by
the 't Hooft limit (where we take large $k$ and look at energies of
order $E \sim M_s / \sqrt{k}$), is still not clear. In principle
correlation functions at these scales may be computed from
supergravity, but so far it is not known if their behaviour (and the
behaviour of the equation of state in this regime) is more like a field
theory (and, if so, which field theory ?) or more like a string
theory.
\item
We have not had a chance to discuss here LSTs in lower dimensions or
with less supersymmetry. It would be interesting to try to classify
these theories and to see if they are similar to the six dimensional
LSTs described above or not. It is also interesting to analyze
compactifications of the LSTs, which give rise to many interesting
theories. Unlike local field theories, the behaviour of non-local field
theories upon compactification is not determined by the behaviour of
the uncompactified theory, so it should be studied independently and
may reveal new results. In particular, the large $k$ behaviour of the
$T^5$ compactification of the LSTs is related to M(atrix) theory on
$T^5$, and it would be interesting to understand it further and to
identify the states whose energy scales as $1/k$ in the large $k$
limit.
\end{enumerate}

It has recently been suggested \cite{gkone,gktwo,gkthree} that the
holographic description of LSTs becomes weakly coupled if one looks at
particular configurations which are far out on the LST moduli
space. For example, configurations in the $(1,1)$ theory where the
low-energy $U(k)$ gauge theory is broken at a scale $M_W$ were argued
to have a perturbation expansion in $M_s/M_W$, which can be used to
reliably compute correlation functions in these theories (far out on
the moduli space) at energies well below $M_W$ (but potentially above
the scales $M_s$ and $M_s/\sqrt{k}$). If perturbation theory is indeed
reliable in these configurations it would be very interesting to use
them for various computations, and to see what they can teach us about
the LSTs.

\section*{Acknowledgments}

I would like to thank T. Banks, M. Berkooz, S. Kachru, D. Kutasov,
N. Seiberg and E. Silverstein for enjoyable collaboration and
discussions on the results presented here.
This research is supported in part by DOE grant DE-FG02-96ER40559.

\end{document}